\documentclass[twocolumn,prb]{revtex4}
\usepackage{amsfonts}
\usepackage{times}
\usepackage{amsmath,amsbsy,amssymb,graphicx}

\begin{document}

\title{Pseudospin 3/2 Fermions, Type-II Weyl Semimetals and Critical Weyl
Semimetals \\
in Tricolor Cubic Lattice}
\author{Motohiko Ezawa}
\affiliation{Department of Applied Physics, University of Tokyo, Hongo 7-3-1, 113-8656,
Japan}

\begin{abstract}
Multi-band touchings such as 3-band, 6-band and 8-band touchings together
with the emergence of high pseudospin fermions were predicted recently at
high-symmetry points in three-dimensional space. In this paper, we propose a
simple cubic model whose unit cell contains three atoms. There are 6 bands
in the system due to the spin degrees of freedom. The 4-band and 2-band
touchings are realized at high-symmetry points, where we derive low-energy
theories, demonstrating the emergence of pseudospin-3/2 fermions and Weyl
fermions, respectively. Away from the high-symmetry points, we find critical
Weyl fermions present exactly at the boundary between the type-I and type-II
Weyl fermions. This critical Weyl fermion transforms into the type-I or
type-II Weyl fermion once the magnetic field is applied.
\end{abstract}

\maketitle

%\address{{\normalsize {Department of Applied Physics, University of Tokyo, Hongo 7-3-1, 113-8656, Japan}}}

%\baselineskip24pt

Dirac, Weyl and Majorana fermions have attracted much attention in condensed
matter physics in views of topology and symmetry. They have a 2-band
touching with a linear dispersion. Dirac and Weyl fermions emerge in various
materials, which are called Dirac and Weyl semimetals\cite{Hosur}. An
interesting feature of Weyl semimetals is that they have a monopole charge
in the momentum space, which protects the existence of the Weyl points
topologically\cite{Murakami}. The type-II Weyl semimetal has attracted much
attention recently\cite{II,Sun, WanMo, Pix,Ob,Ming,Tamai,Tch, Chan,Udagawa},
which emerges when the Weyl cone is highly tilted so that the Fermi surface
consists of a pair of electron- and hole- pockets touching at the Weyl
point. It is experimentally realized in MoTe$_{2}$\cite%
{Huang,Deng,Jing,Liang,Xu}, LaAlGe\cite{Su}, WTe$_{2}$\cite%
{WangW,WuW,Feng,WangW2}, TaIrTe$_{4}$\cite{Khim}, PtTe$_{2}$\cite{Yan} and Ta%
$_{3}$S$_{2}$\cite{Chen}.

Very recently, new types of fermions with multi-band touching were proposed
based on the symmetry analysis as well as the first-principles calculations,
where 3-band\cite{Brad,Chang}, 6-band\cite{Brad} and 8-band\cite{Brad,Wieder}
touchings were reported at high-symmetry points. Especially, it is shown
that the 3-band touchings are well described by the fermions carrying the
pseudospin 1. Furthermore, the 8-band touching has been proposed in
antiperovskites to produce a pseudospin-3/2 fermion\cite{Hsieh}. On the
other hand, the 4-band touching is yet to be realized although some
discussions were given in the supplement of Ref.\cite{Brad}. 
\begin{figure}[t]
\centerline{\includegraphics[width=0.48\textwidth]{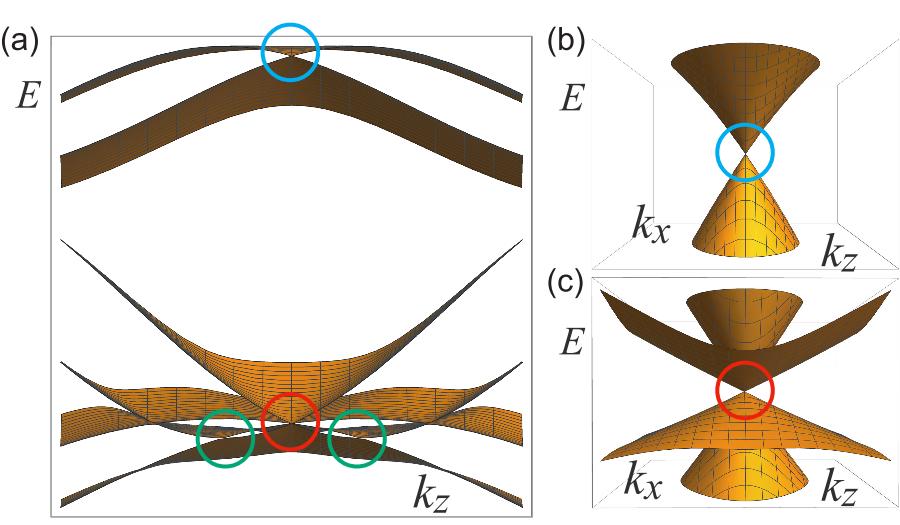}}
\caption{\textbf{Bird's eye's view of the band structure near the $P$ point.}
The horizontal plane is spanned by the $k_{z}$ and $k_{x}$ axes. The
vertical axis is the energy $E=E(k_{x},\protect\pi ,k_{z})$. (a) 2-band
(cyan circle) and 4-band (red circle) touchings are observed at the $P$
point. There appears other 2-band touchings (green circle) off the $P$
point. (b)(c) An enlarged portion of the 2-band (4-band) touching indicates
the emergence of (one) two Weyl cones at the $P$ point. }
\label{FIG1}
\end{figure}

\begin{figure}[t]
\centerline{\includegraphics[width=0.45\textwidth]{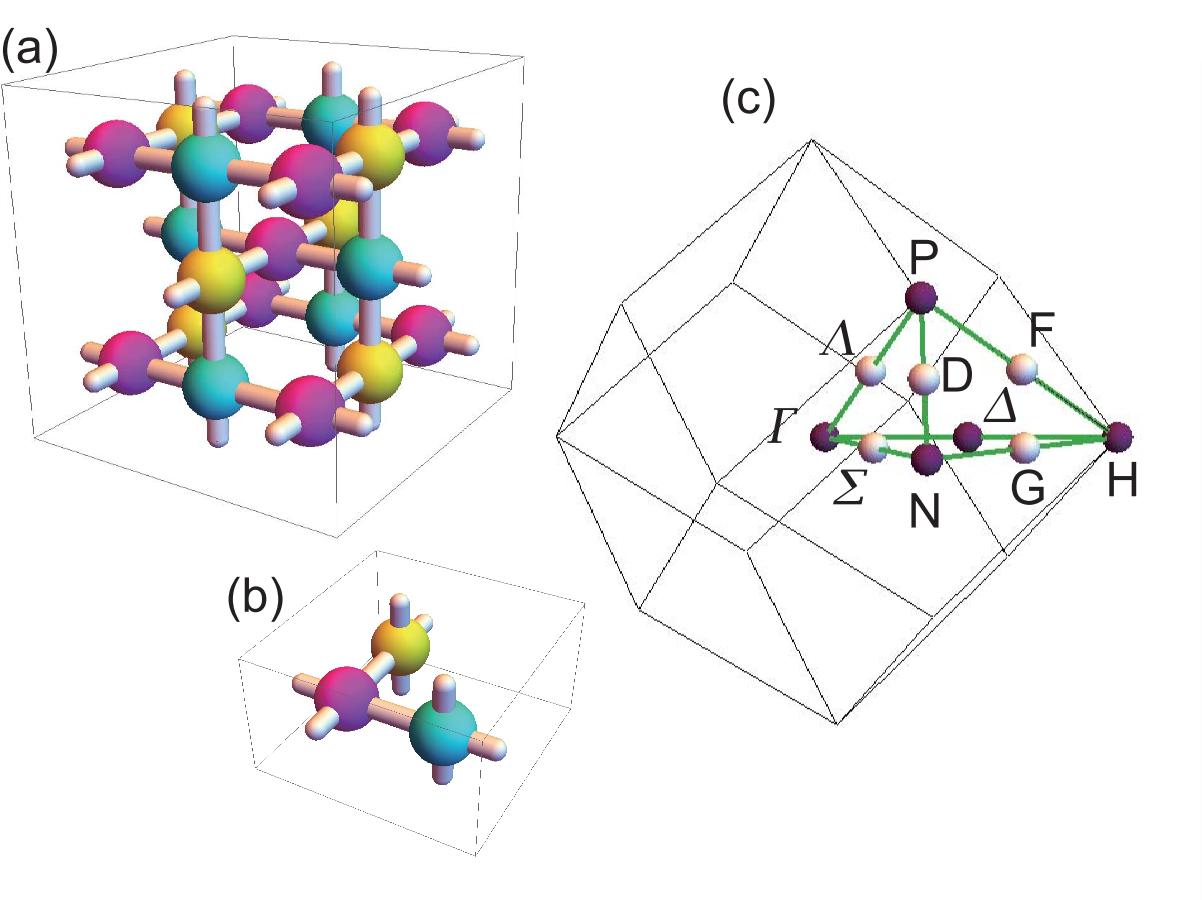}}
\caption{\textbf{Lattice structure and the Brillouin zone.} (a) The lattice
structure of a tricolor cubic lattice. (b) The unit cell contains three
atoms, which are colored by magenta, cyan and yellow. Each colored atom
forms the body-center cubic lattice. (c) The Brillouin zone and the
high-symmetry points $\Gamma $, $P$, $H$, $N$, $\Delta $. The green bold
lines show the cut along the $\Gamma $-$\Delta $-$H$-$G$-$N$-$\Sigma $-$%
\Gamma $-$\Lambda $-$P$-$F$-$H$-$F$-$P$-$D$-$N$ line. }
\label{FIG2}
\end{figure}

In this paper, motivated by these proposals on multi-band touchings and
fermions carrying higher pseudospins, we present tight-binding models
possessing 4-band and 2-band touchings [Fig.\ref{FIG1}]. They are realized
naturally in a lattice structure with the cubic symmetry where the unit cell
contains three atoms [Fig.\ref{FIG2}]. We call it a tricolor cubic lattice.
First, we derive the low-energy 4-band theory at the high-symmetry points,
which is shown to produce pseudospin-3/2 fermions having the angular momenta 
$j=(-3/2,-1/2,1/2,3/2)$. We also show that the bands have monopole charges $%
-3,-1,1,3$ at these points. Second, the 2-band theory is described by the
Weyl fermion. Furthermore, away from the high-symmetry points, we find a
critical Weyl fermion, which resides at the exact boundary of the type-I and
type-II Weyl fermions. They transform into the type-I or type-II Weyl
fermions once the magnetic field is applied.

\textbf{Tricolor cubic lattice:} We consider a cubic lattice as illustrated
in Fig.\ref{FIG2}(a). Though it looks complicated, the unit cell is quite
simple. It contains three atoms represented by magenta, cyan and yellow
spheres [Fig.\ref{FIG2}(b)]. Each color atom forms the body-center cubic
lattice. We call it a tricolor cubic lattice. The Brillouin zone is shown in
Fig.\ref{FIG2}(c). There are four high-symmetry points; $\Gamma \left(
0,0,0\right) $, $P\left( \pi ,\pi ,\pi \right) $, $N\left( \pi ,\pi
,0\right) $ and $H\left( 2\pi ,0,0\right) $. Additionally, there are
important points named $\Delta \left( \pi ,0,0\right) $, $G\left( 3\pi
/2,\pi /2,0\right) $, $\Sigma \left( \pi /2,\pi /2,0\right) $, $\Lambda
\left( \pi /2,\pi /2,\pi /2\right) $, $F\left( 3\pi /2,\pi /2,\pi /2\right) $%
, $D\left( \pi ,\pi ,\pi /2\right) $.

\textbf{Model Hamiltonian:} The main term of the lattice Hamiltonian is the
hopping term along the bonds $\mathbf{d}_{ij}=\mathbf{r}_{i}-\mathbf{r}_{j}$
connecting a pair of the nearest neighbor sites $i$ and $j$ in the tricolor
cubic lattice [Fig.\ref{FIG2}(a)]. We also introduce the spin-orbit
interaction (SOI) preserving the cubic crystalline symmetry\cite{TCImodel},%
\begin{equation}
H_{\text{SO}}=i\lambda \sum_{\left\langle i,j\right\rangle }c_{i}^{\dagger }[%
\mathbf{\sigma }\cdot \mathbf{d}_{ij}]c_{j},  \label{Hamil}
\end{equation}%
with $\lambda $ the coupling strength and $\mathbf{\sigma }=(\sigma
_{x},\sigma _{y},\sigma _{z})$ the Pauli matrix for the spin. The
Hamiltonian has six bands due to the spin degrees of freedom. The 6-band
Hamiltonian reads $\hat{H}_{6}=\sum_{\mathbf{k}}c^{\dagger }(\mathbf{k}%
)H_{6}(\mathbf{k})c(\mathbf{k})$ in the momentum space, where%
\begin{align}
H_{6}(\mathbf{k})& =I_{2}\otimes \left( 
\begin{array}{ccc}
0 & f_{x} & f_{y}^{\ast } \\ 
f_{x}^{\ast } & 0 & f_{z} \\ 
f_{y} & f_{z}^{\ast } & 0%
\end{array}%
\right) +\sigma _{x}\otimes \left( 
\begin{array}{ccc}
0 & g_{x} & 0 \\ 
g_{x}^{\ast } & 0 & 0 \\ 
0 & 0 & 0%
\end{array}%
\right)  \notag \\
& +\sigma _{y}\otimes \left( 
\begin{array}{ccc}
0 & 0 & g_{y}^{\ast } \\ 
0 & 0 & 0 \\ 
g_{y} & 0 & 0%
\end{array}%
\right) +\sigma _{z}\otimes \left( 
\begin{array}{ccc}
0 & 0 & 0 \\ 
0 & 0 & g_{z} \\ 
0 & g_{z}^{\ast } & 0%
\end{array}%
\right) ,  \label{Hamil6}
\end{align}%
with $f_{\alpha }=t\cos k_{\alpha }$, $g_{\alpha }=\lambda \sin k_{\alpha }$%
, $\alpha =x,y,z$, and the $2\times 2$ unit matrix $I_{2}$. We remark that
the SOI is zero ($g_{\alpha }=0$) at the high-symmetry points $\Gamma $, $P$%
, $N$, $H$ and also at the $\Delta $ point.

\begin{figure}[t]
\centerline{\includegraphics[width=0.49\textwidth]{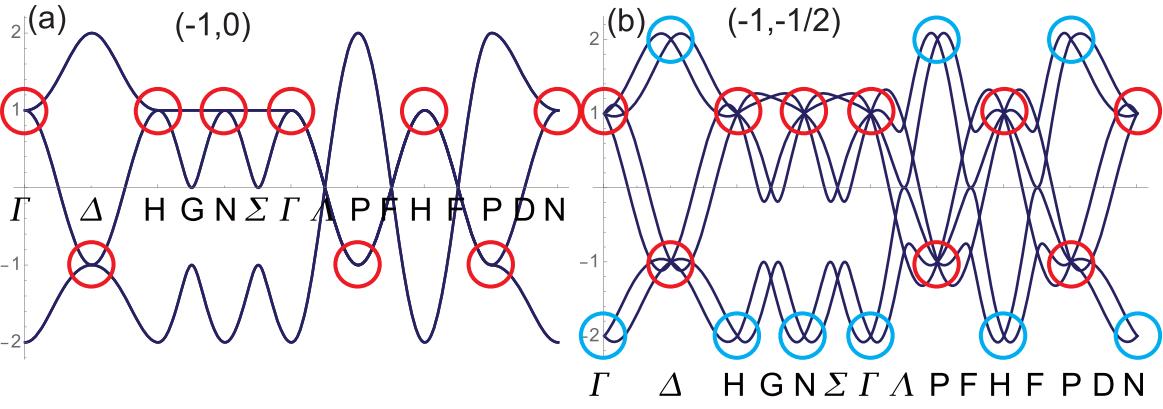}}
\caption{\textbf{Band structures of the 6-band model.} The band structure
along the $\Gamma $-$\Delta $-$H$-$G$-$N$-$\Sigma $-$\Gamma $-$\Lambda $-$P$-%
$F$-$H$-$F$-$P$-$D$-$N$ line for typical values of parameters $(t,\protect%
\lambda)$ as indicated in figures. (a) All energy bands are degenerate with
respect to up and down spins when $\protect\lambda =0$. The 4-fold
degenerate points are marked by magenta circles. (b) The spin degeneracy is
resolved by the SOI ($\protect\lambda \not=0$) except for the high-symmetry
points (indicated by circles) where the SOI vanishes. Pseudospin-3/2
fermions (red) and Weyl fermions (cyan) emerge at various points. }
\label{FIG3}
\end{figure}

\textbf{Band structure:} The energy spectrum is obtained by diagonalizing
the Hamiltonian. We show the band structure along the line $\Gamma $-$\Delta 
$-$H$-$G$-$N$-$\Sigma $-$\Gamma $-$\Lambda $-$P$-$F$-$H$-$F$-$P$-$D$-$N$ in
Fig.\ref{FIG3} for typical values of the parameters $t$ and $\lambda $. The
4-band and 2-band touchings are observed at various points in Fig.\ref{FIG3}%
(b). The high-symmetry point $P$ is typical, around which we show the bird's
eye's view of the band structure in Fig.\ref{FIG1}.

The 4-band touchings are protected by the cubic crystalline symmetry and the
time-reversal symmetry. They occur at the high-symmetry points $\Gamma $, $P$%
, $N$, $H$ and additionally at the $\Delta $ point. Hence, hereafter we
count the point $\Delta $ as a member of the high-symmetry points. Let us
explain how the 4-band touching emerges. As shown in Fig.\ref{FIG3}(a), in
the absence of the SOI, the four-fold degeneracy is present due to the cubic
crystalline symmetry at the high-symmetry points. Even by including the SOI,
the bands never split at the time-reversal invariant momentum points, which
implies the Kramers degeneracy. (In the present model this is realized since
the SOI is zero at these points.) Consequently, the 4-fold degeneracy
without the SOI yields the 4-band touching with the SOI at all the
high-symmetry points [Fig.\ref{FIG3}(b)].

To explore these touchings analytically we diagonalize the Hamiltonian by an
unitary transformation $U$,%
\begin{equation}
U^{-1}H_{6}U=\nu t\text{diag.}(2,2,-1,-1,-1,-1),  \label{diag}
\end{equation}%
at the high-symmetry points; $\nu =+$ for the $\Gamma $, $N$ and $H$ points
and $\nu =-$ for the $P$ and $\Delta $ points. The bands with the first two
energies $2\nu t$ form a 2-band touching, while those with the four energies 
$-\nu t$ form a 4-band touching as in Fig.\ref{FIG3}(b).

\textbf{4-band touching:} First we construct the 4-band model by way of $%
H_{4}(\mathbf{k})=P_{4}U^{-1}H_{6}(\mathbf{k})UP_{4}$, where $U$ is fixed by
(\ref{diag}) while $P_{4}$\ is the projection operator from the $6\times 6$
Hamiltonian to the $4\times 4$ Hamiltonian containing the four bands with
the eigen energies $-\nu t$. The 4-band Hamiltonian is derived up to the
linear order of $k_{\alpha }$ as%
\begin{align}
\nu H_{4}& =-t  \notag \\
& -\frac{\pi \lambda }{3}\left( 
\begin{array}{cccc}
0 & \eta \sqrt{3}k_{z} & 3ik_{y} & \eta \sqrt{3}k_{x} \\ 
\eta \sqrt{3}k_{z} & -2k_{z} & \eta \sqrt{3}k_{x} & 2k_{x}-ik_{y} \\ 
-3ik_{y} & \eta \sqrt{3}k_{x} & 0 & -\eta \sqrt{3}k_{z} \\ 
\eta \sqrt{3}k_{x} & 2k_{x}+ik_{y} & -\eta \sqrt{3}k_{z} & 2k_{z}%
\end{array}%
\right) ,  \label{H4}
\end{align}%
where $\eta =+$ for the $\Gamma $, $P$, $N$, $H$ points and $\eta =-1$ for
the $\Delta $ point. Here we have set the origin of the momentum $k_{\alpha
}=0$ at each high-symmetry point to investigate physics near the point. This
Hamiltonian (\ref{H4}) is exactly diagonalizable,%
\begin{equation}
E_{j}=\nu t\pm \frac{2j}{3}\lambda k,\quad \quad k=|\mathbf{k}|,
\label{Ene4}
\end{equation}%
which is independent of $\eta $,\ where $j=-3/2,-1/2,1/2,3/2$. The energy
spectrum consists of two Weyl cones with velocities $\lambda /3$ and $%
\lambda $ [Fig.\ref{FIG1}(b)].

\textbf{Pseudospin 3/2 fermion:} The energy eigenvalues (\ref{Ene4}) as well
as the eigenfunctions are the same as those of the pseudospin-3/2 Weyl
fermion defined by 
\begin{equation}
H_{4}=\nu t+\frac{2}{3}\lambda \mathbf{k}\cdot \mathbf{J},
\end{equation}%
where $\mathbf{J}$ is the pseudospin-3/2 operator. It implies that the
Hamiltonian (\ref{H4}) is unitary equivalent to the system of the
pseudospin-3/2 fermions.

\begin{figure*}[t]
\centerline{\includegraphics[width=0.98\textwidth]{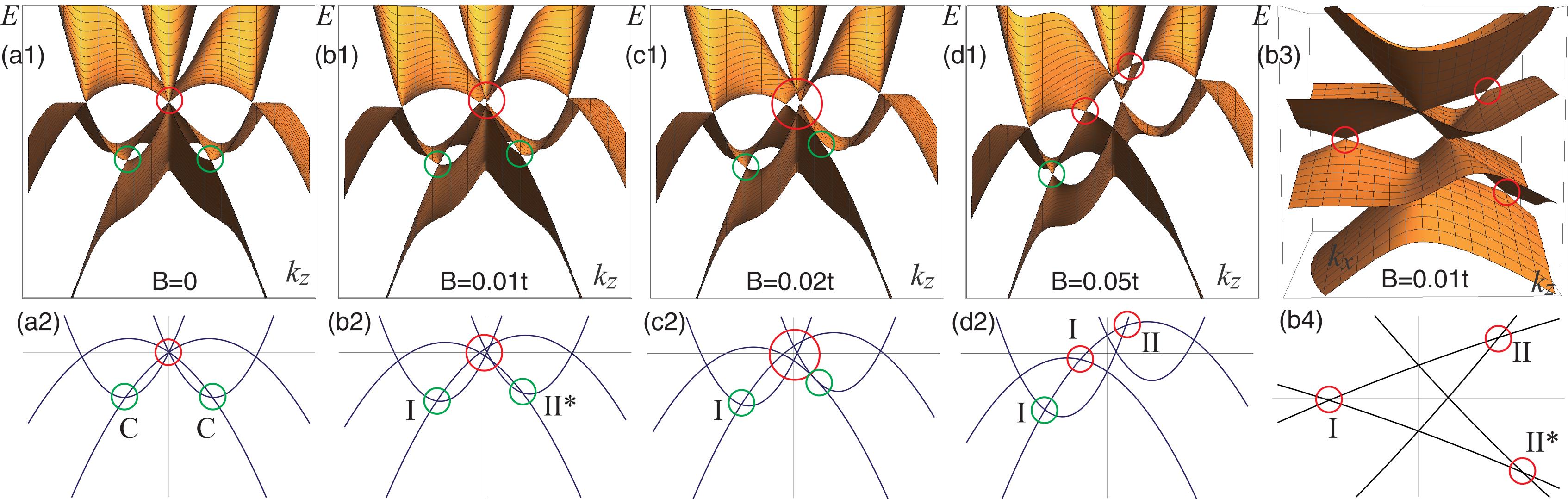}}
\caption{\textbf{Band structures in the vicinity of the $P$ point.} The
horizontal axes are the $k_{z}$ and $k_{x}$ axes, while the vertical axis is
the energy $E$. (a1)$\sim $(d1) The band structure without and with magnetic
field $B$ in the [001] direction. The value of $B$ is given in figures.
4-band touching (red circle) and two 2-band touchings (green) are observed.
(a2)$\sim $(d2) 4-band and 2-band touchings occur along the $k_{z}$ axis ($%
k_{x}=k_{y}=0$). The cross sections of various surfaces are well
approximated by parabolic curves. (b3) and (b4) show enlarged portions of
the vicinity of the 4-band touching in (b1) and (b2), respectively. Symbols
C, I and II stand for the critical, type-I and type-II Weyl points. The
points indicated by II* are identical to those in Fig.\protect\ref{FIG5} and
Fig.\protect\ref{FIG6}. (b2) shows that the 4-band touching (red circle) is
broken but the two 2-band touchings (green) are not. The critical Weyl
points in (a2) turn out to be type-I and type-II Weyl points in (b2). See
also Fig.\protect\ref{FIG5} for details. The pair of the type-II Weyl points
(II*) in (c2) and (b4) are annihilated in (d2). See Fig.\protect\ref{FIG6}
for details.}
\label{FIG4}
\end{figure*}

A comment is in order with respect to the pseudospin-3/2 fermion. It has
already been proposed in antiperovskites, where the Hamiltonian is written as%
\cite{Hsieh,Isobe}%
\begin{equation}
H=m\tau _{z}+v_{1}\tau _{x}\mathbf{k}\cdot \mathbf{J}+v_{2}\tau _{x}\mathbf{k%
}\cdot \mathbf{\tilde{J}}.
\end{equation}%
It is an 8-band model due to the presence of another pseudospin degrees of
freedom $\mathbf{\tau }$. Furthermore, this model has another operator $%
\mathbf{\tilde{J}}$\ preserving the cubic symmetry, which introduces another
velocity $v_{2}$. In general, it is impossible to obtain the exact energy
spectrum of this Hamiltonian.

\textbf{Monopoles:} With the use of the eigen function, the Berry curvature
is explicitly calculated for each band as%
\begin{equation}
\mathbf{\Omega }_{j}=i\nabla \times \left. \left\langle \psi _{j}\right\vert
\nabla \psi _{j}\right\rangle =j\frac{\mathbf{k}}{k},
\end{equation}%
where $j$ labels the band with $j=-3/2,-1/2,1/2,3/2$. In deriving the
formula we have used the fact $\frac{\partial \psi }{\partial r}=0$. Since%
\begin{equation}
\rho _{j}=\frac{1}{2\pi }\iiint \nabla \cdot \mathbf{\Omega }%
_{j}=2j=-3,-1,1,3,
\end{equation}%
the Berry curvature of the band indexed by $j$ describes a monopole with the
monopole charge $2j$.

\textbf{Weyl semimetals:} We may also construct the 2-band model by way of $%
H_{2}(\mathbf{k})=P_{2}U^{-1}H_{6}(\mathbf{k})UP_{2}$, where $P_{2}$ is the
projection operator from the $6\times 6$ Hamiltonian to the $2\times 2$
Hamiltonian containing the two bands with the eigen energies $2\nu t$ in (%
\ref{diag}). The low-energy 2-band model derived from this Hamiltonian
describes Weyl fermions,%
\begin{equation}
\nu H_{2}=2t+\frac{2}{3}\lambda \left( k_{x}\sigma _{x}-k_{y}\sigma
_{y}-k_{z}\sigma _{z}\right) .  \label{Weyl}
\end{equation}%
We may verify the presence of monopole doublets $\pm 1$ at the points $%
\Gamma $, $N$, $H$, and $\mp 1$ at the points $P$, $\Delta $, where the
upper (lower) component dictates the monopole charge of the upper (lower)
band. It is remarkable that Weyl fermions emerge naturally in the present
3-dimensional tight-binding model. The Weyl semimetal is topologically
protected as far as the 2-band touching is intact.

We note that there exist also 2-band touchings at the points $G$ and $\Sigma 
$. However, since the effective theory is derived as%
\begin{equation}
H=\pm \left( -t+\lambda k_{z}\sigma _{z}\right) ,
\end{equation}%
they do not describe Weyl semimetals.

\textbf{Critical Weyl semimetals:} 2-band touchings emerge also at points
which have no conventional names. See points marked by two green circles in
Fig.\ref{FIG1}(a) and Fig.\ref{FIG4}(a1). We study the band structure near
the points. Since they locates on the $k_{z}$ axis, we derive the energy
spectrum of the Hamiltonian $H_{4}$ by setting $k_{x}=k_{y}=0$. It is given
up to the order of $k_{z}^{2}$ as%
\begin{equation}
\nu E(0,0,k_{z})=-t\mp \lambda k_{z}+\frac{t}{2}k_{z}^{2},\quad -t\mp \frac{%
\lambda }{3}k_{z}-\frac{t}{6}k_{z}^{2}.
\end{equation}%
We can check that these parabolic curves fit the results obtained by the 6
band tight-binding model very well. There are 2-band crossing points at $%
k_{z}=\pm \lambda /t$ with the energy $\pm \left( t+\lambda ^{2}/2t\right) $%
: See Fig.\ref{FIG4}(a2). In the vicinity of these points, we derive the
2-band model with the use of $k_{z}^{\prime }=k_{z}\mp \lambda /t$,%
\begin{align}
\nu H& =-t-\frac{\lambda ^{2}}{2t}\mp \frac{1}{3}\lambda k_{z}^{\prime }+%
\frac{t}{3}k_{z}^{\prime 2}  \notag \\
& +\frac{\lambda }{\sqrt{3}}\left( -k_{x}\sigma _{x}+k_{y}\sigma _{y}+\frac{1%
}{\sqrt{3}}k_{z}^{\prime }\sigma _{z}\right) \pm \frac{t}{3}k_{z}^{\prime
2}\sigma _{z}.  \label{Critical}
\end{align}%
The energy spectrum along the $k_{z}$ axis is given by%
\begin{equation}
\nu E(0,0,k_{z}^{\prime })=-t-\frac{\lambda ^{2}}{2t}+\frac{t}{2}%
k_{z}^{\prime 2},\quad -t-\frac{\lambda ^{2}}{2t}\mp \frac{2}{3}\lambda
k_{z}^{\prime }-\frac{t}{6}k_{z}^{\prime 2}.
\end{equation}%
Interestingly, the linear order of $k_{z}^{\prime }$ is absent in the second
energy spectrum. Consequently, they are a critical Weyl semimetals between
the type I and II Weyl semimetals. 
\begin{figure}[t]
\centerline{\includegraphics[width=0.4\textwidth]{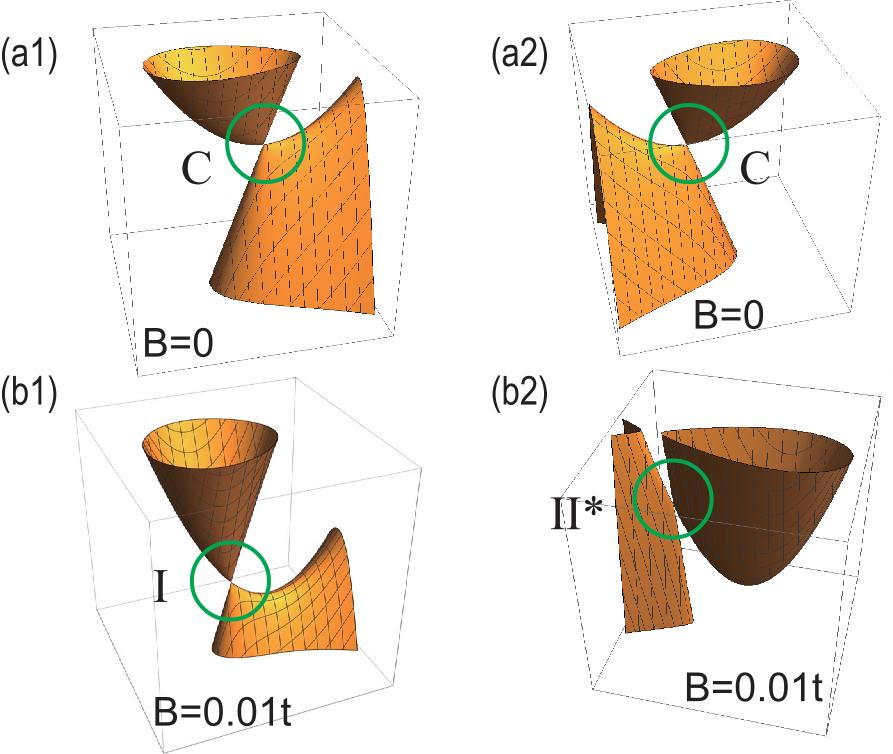}}
\caption{\textbf{Transition from the critical Wey fermions to the type-I and
type-II Weyl fermions.} (a1),(a2) Critical Weyl points. (b1) Type-I Weyl
point. (b2) Type-II Weyl point. They represent bird's eye's views of the
corresponding points in Fig.\protect\ref{FIG4}(a2) and (b2).}
\label{FIG5}
\end{figure}

\textbf{Magnetic-field-induced type-II Weyl semimetals:} We apply external
magnetic field ($B\neq 0$) in the [001] direction. The Hamiltonian is given
by adding the term $B\sigma _{z}\otimes I_{3}$ to the Hamiltonian (\ref%
{Hamil6}), where $I_{3}$ is the $3\times 3$ unit matrix. There occurs the
Zeeman split in the band structure as in Fig.\ref{FIG4}.

With respect to the 2-band touching (Weyl point), the only effect of the
magnetic field is adding the Zeeman term $B\sigma _{z}$ to the Hamiltonian (%
\ref{Weyl}). It results in a shift of the Weyl point in the $z$\ direction,
but the gap never opens.

On the other hand, the 4-band touching is broken under the magnetic field.
This is because it is protected by the time-reversal symmetry and the cubic
symmetry.

\begin{figure}[t]
\centerline{\includegraphics[width=0.48\textwidth]{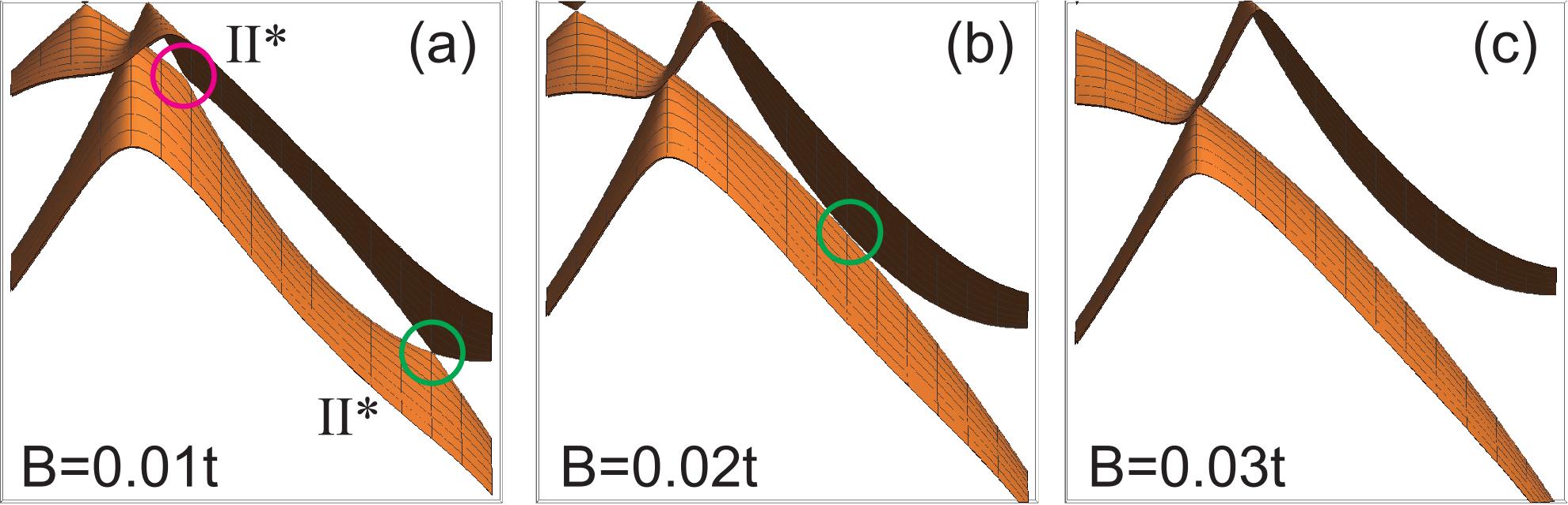}}
\caption{\textbf{Pair-annihilation of two type-II Weyl fermions.} (a) shows
a bird's eye's view of the type-II points marked II* in Fig.\protect\ref%
{FIG4}(b2) and in Fig.\protect\ref{FIG4}(b4). (b) These two Weyl fermions
merge at a critical magnetic field as in Fig.\protect\ref{FIG4}(c2). (c)
They are annihilated beyond the critical field.}
\label{FIG6}
\end{figure}

Type-II Weyl fermions are produced in two different ways as the magnetic
field is introduced. (i) First, as we remarked just in above, there exist
critical Weyl points in the vicinity of the 4-band touching point [Fig.\ref%
{FIG4}(a) and Fig.\ref{FIG5}(a)]. They are turned into the type-I and
type-II Weyl points [\ref{FIG4}(b) and Fig.\ref{FIG5}(b)]. (ii) Second, as
the 4-point touching is broken, there appear six 2-point touchings in
general as in Fig.\ref{FIG4}(b4), among which we find type-II Weyl points.

We discuss the case (i): See Fig.\ref{FIG5}. We study how the critical Weyl
fermions are modified under magnetic field. The crossing points are given by 
$k_{z}=\pm \left( \lambda +\sqrt{\lambda ^{2}+12Bt}\right) /2t$. In the
linear order of $B$, the additional term to the 2-band model (\ref{Critical}%
) is 
\begin{equation}
\nu H_{B}=B\left( \pm 1+\frac{t}{\lambda }k_{z}^{\prime }-\frac{2t}{\lambda }%
k_{z}^{\prime }\sigma _{z}\right) .
\end{equation}%
It transforms the two critical Weyl fermions into the type-I and type-II
Weyl fermions.

Next we discuss the case (ii): See Fig.\ref{FIG4}(b3). In the 4-band linear
model, the energy spectrum along the $k_{z}$ axis is given by $-\nu E=t\pm
\left( \lambda k_{z}-B\right) $, $t\pm \left( \frac{1}{3}\lambda
k_{z}-B\right) $. There are 2-band crossings at $k_{z}=3B/\lambda $. In the
vicinity of this point, the 2-band model is given by%
\begin{equation}
H=\pm \lbrack -t+2B-\frac{2}{3}\lambda k_{z}^{\prime }-\frac{\lambda }{\sqrt{%
3}}(k_{x}\sigma _{x}-k_{y}\sigma _{y}-\frac{1}{\sqrt{3}}k_{z}^{\prime
}\sigma _{z})]
\end{equation}%
with the energy spectrum%
\begin{equation}
E=\pm \left( -t+2B-\frac{2}{3}\lambda k_{z}^{\prime }\pm \frac{1}{3}\lambda 
\sqrt{3k_{x}^{2}+3k_{y}^{2}+k_{z}^{\prime 2}}\right) .
\end{equation}%
Since the tilt of the Weyl cone is larger than the velocity of the Weyl
cone, they are type-II Weyl points.

We have mentioned the emergence of type-II Weyl points in two different
ways. Interestingly they are pair annihilated at $B=\pm \lambda ^{2}/12t$
and disappear for $\left\vert B\right\vert >\lambda ^{2}/12t$, as shown in
Fig.\ref{FIG6}(b) and Fig.\ref{FIG6}(c), respectively.

The author is very much grateful to N. Nagaosa for many helpful discussions
on the subject. He thanks the support by the Grants-in-Aid for Scientific
Research from MEXT KAKENHI (Grant Nos. JP25400317 and JP15H05854).

%\bibitem{} \let\textit{=}\relax

\end{document}